\documentclass[%
reprint,
superscriptaddress,
amsmath,amssymb,
mathtools,
aps, 
pra,
]{revtex4-2}

\usepackage{graphicx}
\usepackage{gensymb}
\usepackage{braket}
\usepackage{dsfont}
\usepackage{comment}
\usepackage{siunitx}
\usepackage[colorlinks=true,allcolors=blue]{hyperref}
\usepackage[capitalise]{cleveref}
\usepackage{layouts}
\usepackage{upgreek,textcomp}

\newcommand{\eref}[1]{Eq.~(\ref{#1})}
\newcommand{\fref}[1]{Fig.~\ref{#1}}
\newcommand{\tref}[1]{Table~\ref{#1}}

\newcommand{\aref}[1]{App.~\ref{#1}}

\DeclareSIUnit{\nothing}{\relax}

\begin{document}

\title{Modification of adhesion between microparticles and engineered silicon surfaces}

\author{Fabian Resare}
\affiliation{Department of Microtechnology and Nanoscience (MC2), Chalmers University of Technology, SE-412 96 G\"oteborg, Sweden}
\author{Somiya Islam Soke}
\affiliation{Department of Microtechnology and Nanoscience (MC2), Chalmers University of Technology, SE-412 96 G\"oteborg, Sweden}
\author{Witlef Wieczorek}
\email{witlef.wieczorek@chalmers.se}
\affiliation{Department of Microtechnology and Nanoscience (MC2), Chalmers University of Technology, SE-412 96 G\"oteborg, Sweden}


\begin{abstract}
    A key challenge in performing experiments with microparticles is controlling their adhesion to substrates. For example, levitation of a microparticle initially resting on a surface requires overcoming the surface adhesion forces to deliver the microparticle into a mechanical potential acting as a trap. By engineering the surface of silicon substrates, we aim to decrease the adhesion force between a metallic microparticle and the silicon surface. To this end, we investigate different methods of surface engineering that are based on chemical, physical, or physio-chemical modifications of the surface of silicon. We give quantitative results on the detachment force, finding a correlation between the water contact angle and the mean detachment force, indicating that hydrophobic surfaces are desired for low microparticle adhesion. We develop surface preparations decreasing the mean detachment force by more than a factor of three compared to an untreated silicon surface. Our results will enable reliable levitation of microparticles and are relevant for experiments requiring low adhesion between microparticles and a surface.
\end{abstract}

\maketitle

\section{Introduction} 

Levitation of microscopic objects has become a useful tool for studies in materials science \cite{Ge2018density, Ge2020MagLev, Voisin24_PRL} as well as in applied or fundamental physics experiments \cite{Romero-Isart2011, Millen2020_review, Gonzalez-Ballestro2021}. In particular, magnetic levitation of superconducting microparticles is an emerging levitation method, predicted to lead to a new class of fundamental physics experiments \cite{Romero-Isart2012_QuaMag, Cirio12, QST_Pino2018,carney2024_dm_maglev, Higgins24_DM}, as well as applications in force or acceleration sensing \cite{Johnsson2016_gravimetry,Prat-Camps2017_sensing}, with recent experimental demonstrations \cite{waarde_thesis, marti_pra, vienna_levitation}. 

A basic challenge in levitation of microscopic or even smaller solid-state objects is to load the object into the potential trap. Magnetic trapping of superconducting particles requires operation in cryostats providing the necessary low temperature and vacuum environment. As levitation can only occur when the particle is superconducting, loading mechanisms that have been used in optical or electrical levitation experiments are generally unsuitable. Typically, particle sizes in these experiments are sub-\textmu m, while magnetic traps for superconductors use sizes of tens of \textmu m \cite{marti_ieee,marti_pra,vienna_levitation}. Reported loading mechanisms for nanoparticles either disperse particles using light \cite{northup_liad, Rieser24}, which breaks superconductivity, rely on piezoelectrically induced shaking \cite{barker_piezo, winstone_apparatus, Wang24, Khodaee22}, which causes too much heat load at low temperatures, use aerosols \cite{kieselCavityCoolingOptically2013,Piotrowski23}, which would freeze out at low temperatures, or deliver particles into the trap via a low-vacuum room temperature environment \cite{Piotrowski23, hollow_core_load}, whereby particles would not be superconducting. In case of \textmu m-sized particles, piezoelectric force generation \cite{Moore14} and sublimation-activated release have been successfully used \cite{Murphy24}, but both methods are unsuited for low temperature loading. Hence, in the case of a superconducting magnetic trap, the most straight-forward loading strategy is to place the micrometer-sized object at a position from which it could levitate by means of the magnetic lift force alone, provided the adhesion and gravitational forces are lower.

In our work, we investigate methods to engineer the adhesion force between a spherical microparticle and a silicon substrate. We implement physical, physio-chemical, and chemical surface modifications of a silicon substrate and study their effect on the adhesion of \SI{50}{\micro\meter} diameter Sn$_{63}$Pb$_{37}$ spherical particles with a mass of \SI{0.5}{\micro\gram}. Such particles become superconducting below \SI{6.4}{\kelvin} \cite{marti_ieee} and have been used in magnetic levitation experiments previously \cite{marti_ieee,marti_pra}. Crucially, we determine the detachment force between the particles and different surfaces experimentally. This allows us to show quantitatively that we can reduce the mean detachment force  by at least a factor of three when using a PTFE membrane on Si. This reduction of adhesion will enable reproducible magnetic levitation of smaller particles \cite{Romero-Isart2011,QST_Pino2018} and can be directly transferred to other experiments, where adhesion of microparticles resting on surfaces should be minimized \cite{Ge2018density, Ge2020MagLev, Voisin24_PRL}.

\begin{figure}[b!htp]
    \centering
    \includegraphics[width=\linewidth]{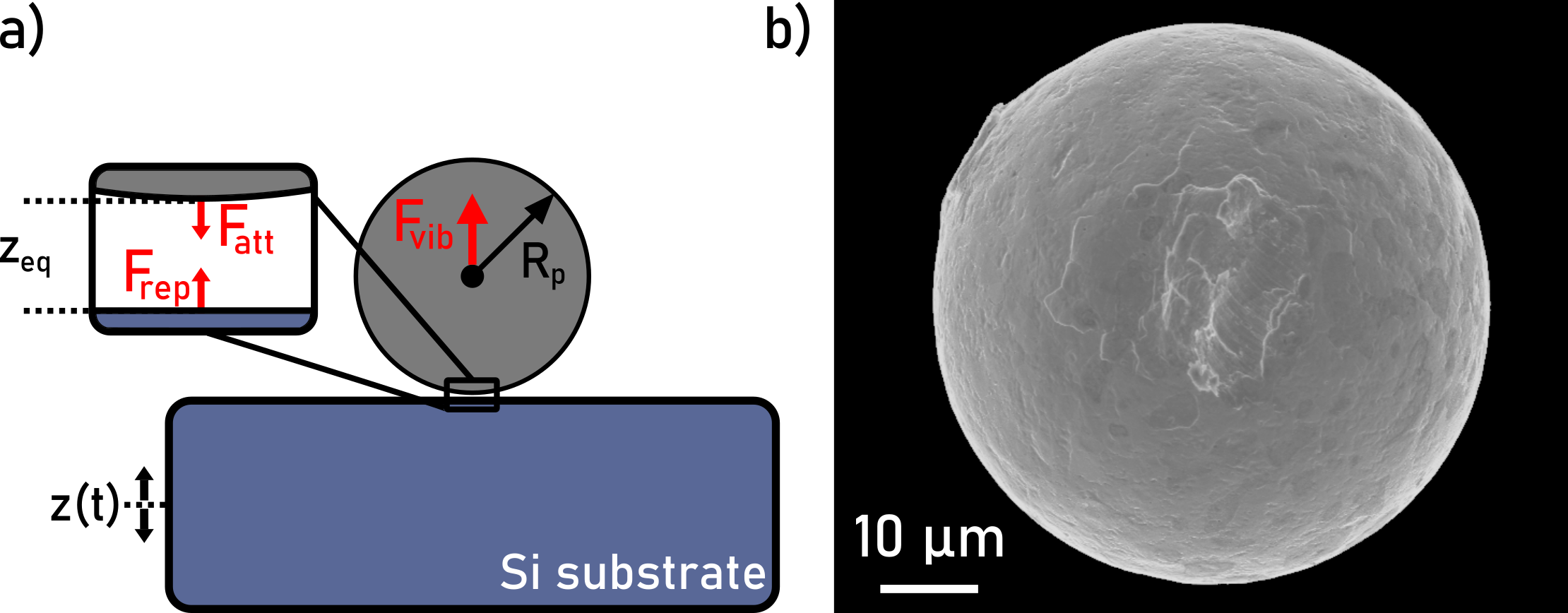}
    \caption{(a) Schematic of the experiment to determine the adhesion of microparticles on a Si surface. At a vertical distance $z_{\text{eq}}$, attractive ($F_\text{att}$) and repulsive forces ($F_\text{rep}$) are in equilibrium for a particle of radius $R_\text{p}$. An oscillatory motion $z(t)$ is applied to the microparticle by shaking the substrate using a surface transducer imparting an upwards force $F_{\text{vib}}$ onto the particle. (b) SEM image of a \SI{50}{\micro\meter} diameter Sn$_{63}$Pb$_{37}$ spherical microparticle.}
    \label{fig:particle_on_surface}
\end{figure}

\section{Theoretical background}

We first discuss relevant forces that act between a surface and a particle. We consider a case like that shown in \fref{fig:particle_on_surface}(a), where a spherical particle is initially positioned at its equilibrium distance $z_{\text{eq}}$ on top of a flat substrate. \fref{fig:particle_on_surface}(b) shows a scanning electron microscope (SEM) image of an actual microparticle, which is highly spherical.

A commonly used model for the force between interacting surfaces is given by the Lennard-Jones (LJ) potential \cite{lj_pot, lj_adhesion}, which in principle is agnostic to the exact physiochemical adhesion mechanisms. The LJ model yields for the maximum adhesion force (for details see \aref{app:forces})
$ F  =  0.497A_{\text{H}, ij}R_{\text{p}}/z_{\text{eq}}$, where $R_{\text{p}}$ is the particle radius, $A_{\text{H}, ij}$ is the material-specific Hamaker constant between surface $i$ and $j$, and $z_{\text{eq}}$ is the equilibrium distance between the two surfaces. The Hamaker constant is a measure of the relative surface free energy between the materials and given through the combining relation $A_{\text{H}, ij}\propto\sqrt{\gamma_{\text{S}, i}\gamma_{\text{S}, j}}$ \cite{Padday68_adhesion,israelachviliIntermolecularSurfaceForces2011}, where  $\gamma_{\text{S}, k}$ is the surface free energy of surface $k$. From this relation it follows that the force required for particle detachment can be decreased by decreasing the surface free energy for either of the surfaces. To get an estimate of the expected adhesion, we assume a typical Hamaker constant of $A_\text{H}\sim\SI{1e-19}{\joule}$ \cite{Butt05_AFM} and particle-surface distance of $z_{\text{eq}}\sim\SI{0.5}{\nano\meter}$ to obtain an adhesion force of $F\sim \SI{5}{\micro\newton}$. Importantly, the LJ model neglects several factors that also determine adhesion, such as surface roughness, presence of liquids, or the real geometry of the contact area. Adhesion can additionally be caused by Coulomb or capillary forces. The Coulomb force between a charged conducting particle and surface can be mitigated by keeping both the particle and the substrate grounded \cite{Vallabh2014_Charge}. Capillary forces are a result from moisture forming a meniscus between the particle and substrate \cite{Grobelny06}. They can be reduced by using solvents with high vapor pressure, allowing time to dry after particle dispersal, or a vacuum bake-out. Finally, it should be noted that the gravitational force of a \SI{50}{\micro\meter} diameter Sn$_{63}$Pb$_{37}$ particle of about \SI{5}{\nano\newton} is negligible. Previous work determined detachment forces in the range of 20-\SI{100}{\nano\newton} for $\SI{2.5}{\micro\meter}$ spherical Au particles on a Si surface \cite{Bohme73}. Using these results, we expect an adhesion force in the range of 400-\SI{2000}{\nano\newton} for the $\SI{50}{\micro\meter}$ diameter microspheres we use. Due to the complexity of the mechanisms causing adhesion, we will quantitatively determine the adhesion force in our work. 

The adhesion force must be overcome by a suitable lift force to achieve release of a particle from the surface. As an example of such a force, we consider a magnetic lift force that is acting on a superconducting particle, proportional to $(\mathbf{B}\cdot\nabla)\mathbf{B}$, where $\mathbf{B}$ is the magnetic field \cite{marti_ieee, marti_pra}. For a single circular coil of radius $R_{\text{coil}}$, the maximal lift force is bounded by the critical field on the particle's surface to $F_{\text{lift}}=\pi B_C^2R_{p}^3/\mu_0 R_{\text{coil}}\simeq\SI{800}{\nano\newton}$, assuming a coil with $R_{\text{coil}}=\SI{300}{\micro\meter}$ and $B_C=\SI{80}{\milli\tesla}$ of a pure Pb particle. This situation is similar to the chip-based traps of Refs.~\cite{marti_ieee,marti_pra}, where superconducting microparticles initially rest on Si chips. With these traps, we find that a lift force of 90 to 700\,nN$\cdot\left(I/[1\text{A}]\right)^2$ can be generated depending on the initial placement of the particle on the chip substrate for a current $I$ through the trap coils (for details see \aref{app:forces}). Overall, the magnetic lift force and the expected detachment force cover a similar force range of between 100 to \SI{1000}{\nano\newton}. Thus, finding a method to reduce adhesion is highly desired for successful levitation experiments.

\begin{figure}[b!htp]
    \centering
    \includegraphics[width=\linewidth]{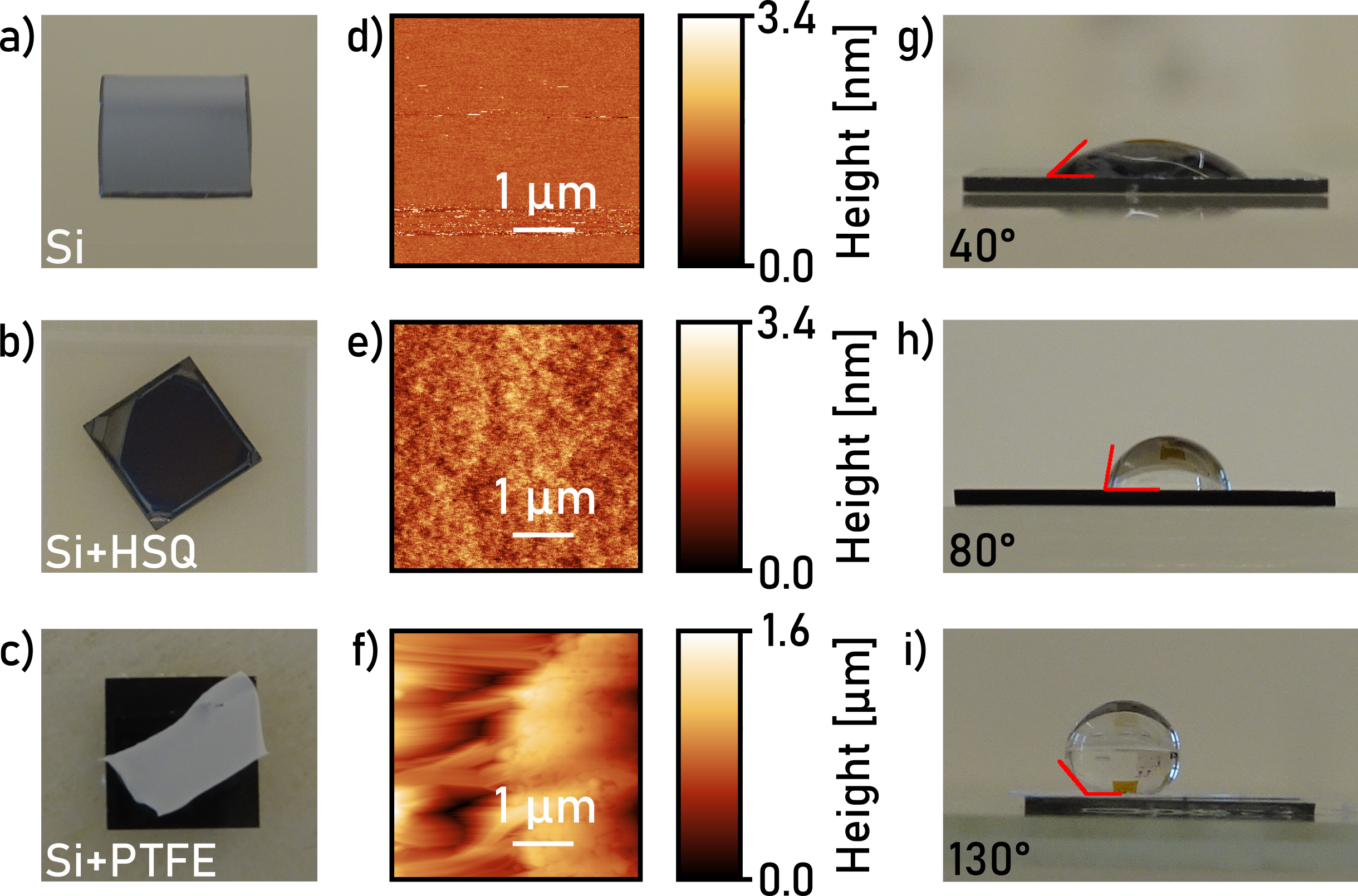}
    \caption{Sample preparation and analysis for selected surface modifications. (a,b,c) photographs of samples, (d,e,f) height scans measured by AFM, and (g,h,i) water droplets on samples to determine the water contact angle (WCA).}
    \label{fig:surface_morphology}
\end{figure}

\section{Experimental measurements}

\subsection{Surface modifications and analysis}

In the following, we present our results on the effect of various physical, chemical, and physio-chemical surface modifications of a Si surface with the goal to alter the adhesion force between this modified surface and the particle. Our modifications are guided by noting that different surface chemistries have different surface-free energies \cite{Kinloch12_adhesion}, that surface roughness plays a crucial role in the strength of the adhesion forces \cite{adhesion_microelectronics}, and that any modification must be high vacuum and cryo-compatible for the application we envision.

In the study, we consider modifications of the surface of the Si substrates only, while we leave the surface of the particle pristine (see \aref{app:sample_prep} for particle preparation), which, however, could be modified as well. By not modifying the particles, we aim to provide a broad method to reduce adhesion for any metallic microparticle.

We used \SI{7}{\milli\meter}$\times$\SI{7}{\milli\meter} Si chips (see \fref{fig:surface_morphology}(a)) that were cut from \SI{280}{\micro\meter} thick $\langle 100\rangle$-oriented undoped silicon wafers as base substrate. As reference samples, we used (1) untreated Si. For purely physical surface modifications we performed (2) microstructuring by potassium hydroxide (KOH) etching and (3) nanostructuring by plasma etching, with the goal of decreasing the number of physical contact points between particle and surface.

As chemical treatments, we employed (4) parylene C chemical vapour deposition including plasma processing, (5) poly methyl methacrylate (PMMA) spin coating, (6) hydrogen silsesquioxane (HSQ) spin coating  (see \fref{fig:surface_morphology}(b)), and (7) gold (Au) physical vapour deposition. The chemical treatments primarily result in a change of the surface-free energy. As physio-chemical treatment we use (8) polytetrafluoroethylene (PTFE) membranes  (see \fref{fig:surface_morphology}(c)), see also \cite{Khodaee22}. Further details on the sample preparations can be found in \aref{app:sample_prep}.

\subsection{Surface morphology}

We analyzed the surface morphology using a combination of SEM and atomic force microscopy (AFM), see \fref{fig:surface_morphology}(d,e,f), \aref{app:sample_prep}, and \aref{app:data}. We find that the physical surface modifications succeeded in achieving a highly structured surface, with the plasma etching resulting in unordered \SI{}{\nano\meter}-scale and the KOH etching in ordered \SI{}{\micro\meter}-scale structuring. 
The chemical treatments using PMMA and HSQ (see \fref{fig:surface_morphology}(e)) resulted in a surface morphology that is practically unchanged to untreated silicon. The physio-chemical treatment with a PTFE membrane shows a highly structured surface on both the \SI{}{\micro\meter}- and \SI{}{\nano\meter}-scales (see \fref{fig:surface_morphology}(f)), with a preferred ridge direction (see SEM in \aref{app:sample_prep}). 

The AFM was operated in peak force mode, yielding not only information on the surface structure, but also a value of the adhesion force $F_{\text{adh}}^{\text{AFM}}$ between the AFM cantilever tip and the particle. \tref{tab:stats} shows the measured values averaged over the scan area, together with their root-mean-square deviation. Because of the difference in size and material of the tip and the microparticles, $F_{\text{adh}}^{\text{AFM}}$ is not sufficient to quantitatively predict particle detachment forces. However, we observe a correlation between our determined detachment force and $F_{\text{adh}}^{\text{AFM}}$, see \tref{tab:stats} and \fref{fig:correlations}.

\subsection{Water contact angle measurement}

We analyzed the wettability of the surfaces by measurement of the water contact angle (WCA) of a water droplet applied on the surface using a pipette, see \fref{fig:surface_morphology}(g,h,i) and \aref{app:data}. This technique provides an estimate of the surface-free energy \cite{israelachviliIntermolecularSurfaceForces2011,Kinloch12_adhesion}, but is insufficient to give a quantitative measure on the particle adhesion as different surface interactions are involved in wetting \cite{degennes_wetting} and microparticle adhesion. We find significant variation of the WCA between the prepared surfaces consistent with substantial changes in chemical and physical surface properties, see \tref{tab:stats}. In particular, we find that the purely physical structuring makes the WCA smaller, in agreement with the Wenzel model \cite{Wenzel1936}, while some chemical treatments increase the WCA, see, e.g., \fref{fig:surface_morphology}(h) and \fref{fig:surface_morphology}(i) for the treatments with HSQ and PTFE, respectively. Literature values of the surface free energy $\gamma_S$ (see \tref{tab:stats}) show a similar spread as our measured WCAs and are inversely related to the WCA results.

\begin{figure}[t!hbp]
    \centering
    \includegraphics[width=\linewidth]{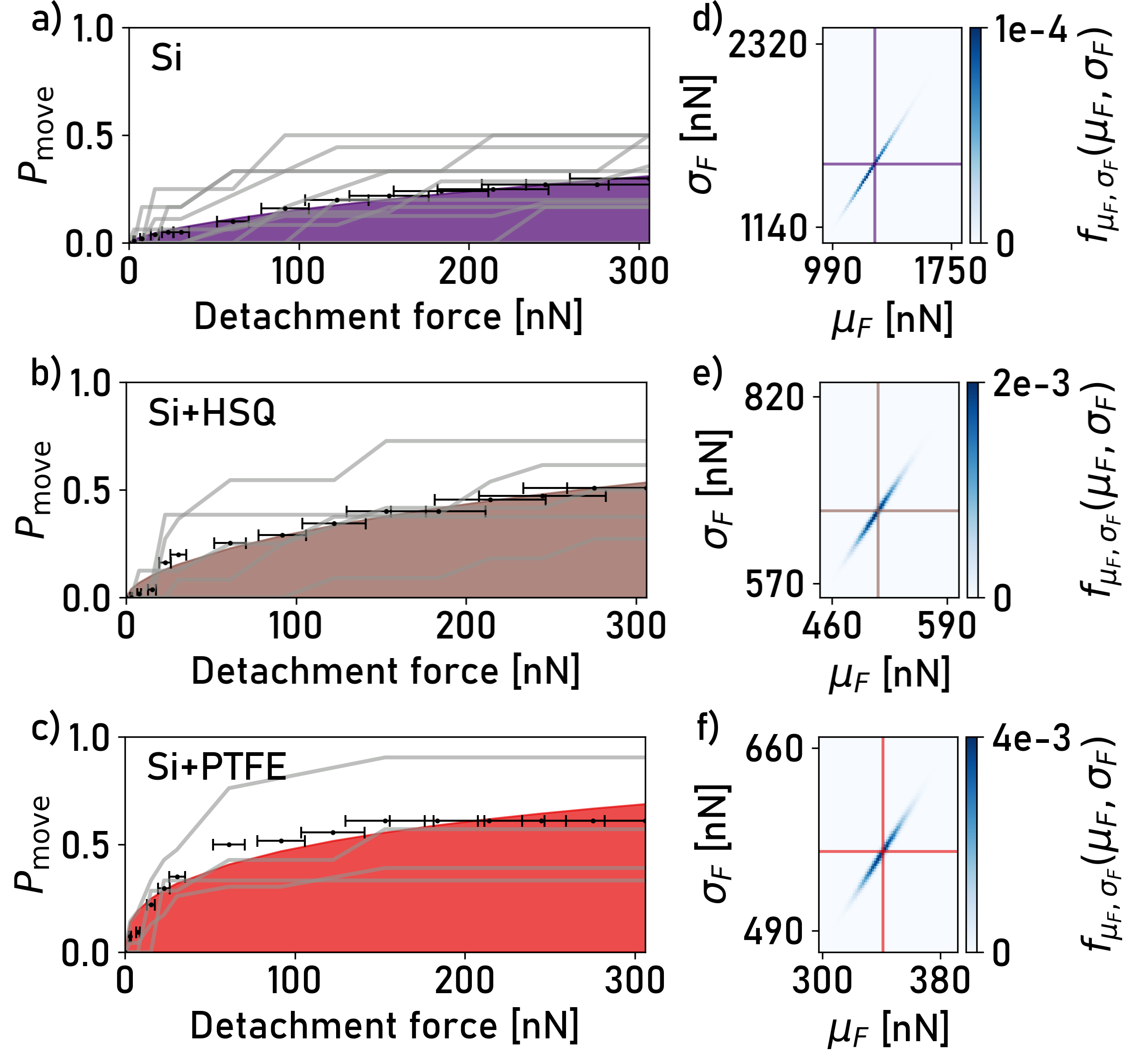}
    \caption{Experimental determination of detachment force. Cumulative distribution functions for (a) the reference Si surface, (b) the HSQ treated surface, and (c) the surface with PTFE. The gray lines show the results of individual samples, while the black line shows the average of all samples. The colored shaded area depicts the fitted gamma distribution model. (d,e,f) show the mean and standard deviation of the posterior probability distributions for the detachment force obtained from MCMC sampling, with the cross marking the 50th percentile.}
    \label{fig:si_cdf}
\end{figure}

\subsection{Detachment force measurement}

\begin{table*}[b!htp]
    \centering
    \begin{tabular}{lcccccccc}\hline\hline
         No.&{Sample} & $\mu_F$ [nN] & $\sigma_{F}$ [nN] & $F_{P=0.5}$ [nN] & $S_q$ [nm] & {WCA [$\degree{}$] ($\pm 10\degree{}$)} & $F_{\text{adh}}^{\text{AFM}}$ [nN] & $\gamma_\text{S}$ [mJ/m$^2$] \\\hline\hline
         \multicolumn{7}{l}{\textbf{No surface modification}}\\\hline
         (1)& Si & $1250^{+360}_{-303}$ & $1530^{+490}_{-400}$ & $700^{+100}_{-100}$ & 0.2 & 40 & 4.7$\pm$0.3 & 2130 \cite{Jaccodine63_SiSurfaceEnergy}\\\hline
         \multicolumn{7}{l}{\textbf{Physical surface modification}}\\\hline
         (2)& Si+KOH etching &  $3720^{+1400}_{-1060}$ & $5770^{+2350}_{-1740}$ & $1410^{+110}_{-130}$ & n/a & 35 & n/a & n/a \\\hline
         (3)& Si+plasma etching & $3150^{+1080}_{-860}$ & $5650^{+2070}_{-1608}$ & $810^{+100}_{-110}$ & 126 & wetting & 3.3$\pm$3.7 & n/a \\\hline
         \multicolumn{7}{l}{\textbf{Chemical surface modification}}\\\hline
         (4)& Si+Parylene C+Plasma & $730^{+180}_{-160}$ & $800^{+220}_{-190}$ & $470^{+70}_{-80}$ & 3 & 110 & 0.9$\pm$0.1  & 40  (untreated) \cite{Golda-Cepa14_parylene}\\\hline
         (5)& Si+PMMA & $1000^{+320}_{-260}$ & $860^{+330}_{-250}$ & $770^{+80}_{-90}$ & 0.4 & 85 & 1.9$\pm$0.2 & 40 \cite{Kinloch12_adhesion}\\\hline
         (6)& Si+HSQ & $510^{+120}_{-110}$ & $670^{+160}_{-150}$ & $270^{+50}_{-50}$ & 0.4 & 80 & 0.9$\pm$0.2  & n/a \\\hline
         (7)& Si+Au & $4900^{+1900}_{-1450}$ & $8780^{+3750}_{-2740}$ & $1230^{+110}_{-130}$ & 1 & 55 & 3.9$\pm$0.9 & 750 \cite{Tran2016_surface_energy} \\\hline
         \multicolumn{7}{l}{\textbf{Physio-chemical surface modification}}\\\hline
         (8)& Si+PTFE & $340^{+80}_{-70}$ & $560^{+140}_{-120}$ & $110^{+20}_{-20}$ & 87 & 130 & 0.5$\pm$0.2 & 20 \cite{Kinloch12_adhesion} \\\hline\hline
    \end{tabular}
    \caption{Adhesion results for different samples: mean detachment force $\mu_F$, standard deviation $\sigma_F$ of the detachment force and the force at which 50\% of particles are expected to move including error intervals, root-mean-square roughness $S_q$ determined from AFM, measured water contact angle (WCA), adhesion force $F_{\text{adh}}^{\text{AFM}}$ measured by AFM, and literature values for the surface free energy $\gamma_\text{S}$. Note that the error intervals in the force values include the uncertainty due to drift in the calibration.}
    \label{tab:stats}
\end{table*}

We determine the detachment force quantitatively by using a calibrated surface transducer to shake off particles from the surfaces under test (details see \aref{app:setup}). In a standard laboratory environment (temperature around 20$\,^\circ$C at unregulated humidity), we deterministically place between 10 to 20 particles on each chip using a micromanipulator needle. The needle is grounded to avoid charging of the particles. Note that the substrate chips may become charged through triboelectricity, with the surface modification affecting the charging dynamics. For example, PTFE is prone to negative charging \cite{triboelectric_series}, which could in principle lead to both an increase or a decrease of adhesion. In the presented experiments, no treatments were done to discharge the substrate surfaces.

The chip is then placed on a surface transducer, driven by an audio amplifier to which we apply a sinusoidal voltage with amplitude $V_0$ to create an oscillatory motion $z(V_0, \omega)=A_{0}(V_0,\omega)\cos{(\omega t)}$ with frequency $\omega$. We independently calibrated the motional amplitude $A_{0}$ using an accelerometer (see \aref{app:setup}).  We worked at a frequency of $\omega/2\pi=453$\,Hz, slow enough to be within the bandwidth of both the accelerometer and amplifier. Although saturation of the amplifier limited the achievable acceleration, we could reach the sensing range limit of the accelerometer of $\pm64g$. All measurement results that we present are within the calibrated range. Particles will detach from the chip surface once the acceleration force $m\omega^2A_{0}$ is larger than the detachment force, with $m$ being the particle mass. In the experiment, we increase $A_0$ in steps and determine the number of particles that moved at each step using a motion detection algorithm (details in \aref{app:setup}). 

\fref{fig:si_cdf}(a,b,c) show results of these measurements for untreated Si, HSQ on Si, and PTFE on Si (for the other remaining surfaces see \aref{app:data}). We show the cumulative distribution function for the probability of particles that moved at a specific detachment force. Note that we can apply a maximal force of \SI{306}{\nano\newton}. \fref{fig:si_cdf}(a) shows the result for the untreated Si surface on which 100 particles were tested. We principally observe that most particles do not detach even at the maximum force applied. However, we observe a considerable decrease in adhesion for the HSQ and PTFE treated samples, see \fref{fig:si_cdf}(b) and \fref{fig:si_cdf}(c), respectively, which were tested with about 50 particles each. In both cases, the measured detachment probability at maximal applied force is nearly a factor of two larger than for the untreated Si sample.

To make a more quantitative analysis beyond the applicable maximal force, we fit a gamma distribution to the data for obtaining the expected mean detachment force and its standard deviation. We motivate this model primarily due to its closeness to the experimental data. Using Markov chain Monte Carlo (MCMC) sampling of parameters for the gamma distribution model, we obtained probability histograms for the mean ($\mu_F$) and standard deviation ($\sigma_F$) of the detachment force, see \fref{fig:si_cdf}(d,e,f). Based on these probability distributions, we can ascertain their mean values and uncertainty intervals between the 16th and 84th percentiles. We can also predict the force at which 50\% of particles are expected to move, $F_{P=0.5}$. \tref{tab:stats} shows the values for all estimated quantities. Further details on the parameter estimation are available in \aref{app:data}. Note that for mean values larger than about \SI{1000}{\nano\newton} the probability distributions are rather broad and, thus, the uncertainty intervals are rather large. This is due to the fact that we could only record data up to \SI{306}{\nano\newton}, which limits the predictive capability of the model for larger detachment forces. Overall, the PTFE coating provided the smallest mean adhesion force, followed by the HSQ coating. The former has, however, the disadvantages of being non-transparent and of adding a thickness of \SI{75}{\micro\meter} to the substrate, which may not be acceptable for certain applications.

\begin{figure}[t!hbp]
    \centering
    \includegraphics[width=\linewidth]{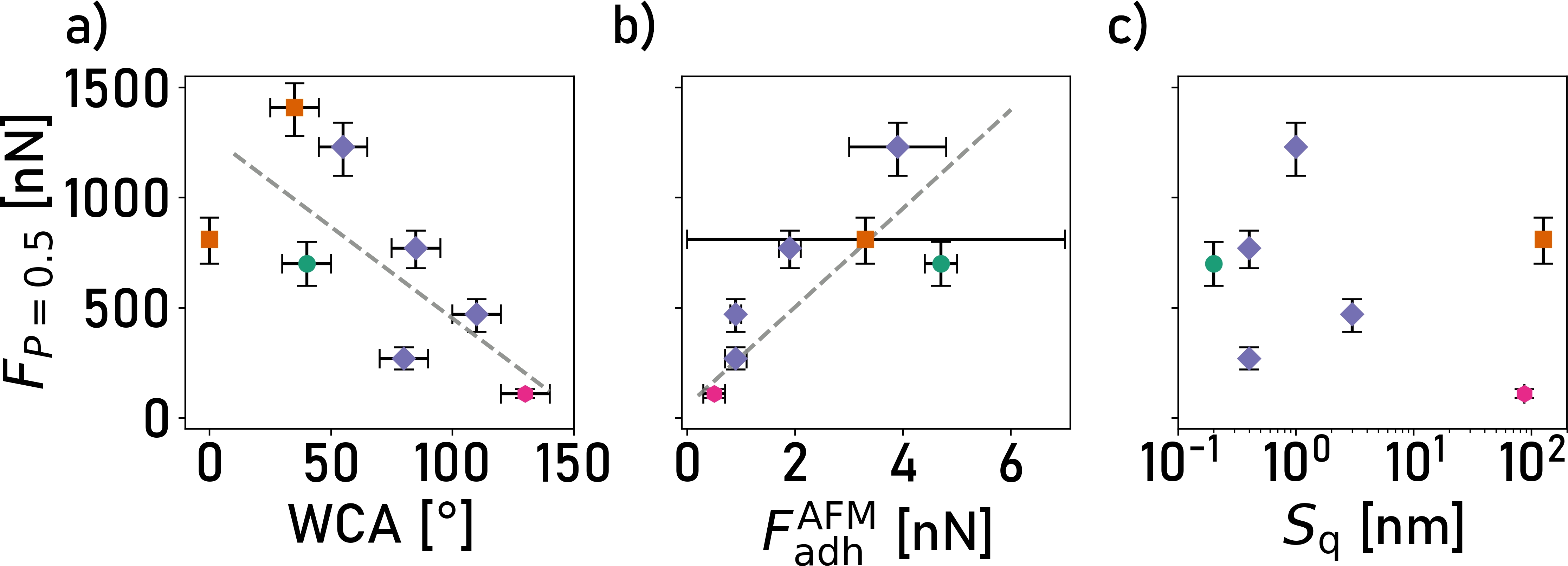}
    \caption{Correlations between the 50\%-detachment force and (a) the water contact angle, (b) the adhesion force measured by the AFM $F_{\text{adh}}^{\text{AFM}}$, and (c) the mean squared roughness $S_{\text{q}}$. We show the data for the untreated sample (green circle), the physical treatments (orange square), the chemical treatments (purple diamond), and the physio-chemical treatment (magenta hexagon). The dashed lines serve as guides to the eye for indicating trends.}
    \label{fig:correlations}
\end{figure}

Finally, to illustrate a possible correlation between measured surface properties and the detachment force, we provide a graphical representation of the data from \tref{tab:stats} in \fref{fig:correlations}. We observe a correlation between small detachment force and high water contact angle (\fref{fig:correlations}(a)) as well as between small detachment force and low adhesion force measured by the AFM (\fref{fig:correlations}(b)). However, with regards to surface roughness (\fref{fig:correlations}(c)), no clear trend is observable. The surface roughness was practically always varied together with the chemical composition of the surface and no correlation seems to be evident.

\section{Conclusion}

We have presented various methods for decreasing adhesive forces between Si substrates and microparticles, with the most successful ones exploiting HSQ coating or PTFE membranes. These resulted in a reduction in mean detachment force by at least a factor of three compared to an untreated Si surface, or a factor of six in the detachment force at which 50\% of particles move. We found a correlation between the hydrophobicity of a surface measured through the water contact angle and the detachment force of the particle, which can be connected to the surface-free energy of the samples \cite{israelachviliIntermolecularSurfaceForces2011,Kinloch12_adhesion}. This correlation can be used as an indicative method for a simple identification of promising surface treatments. We see applicability of our results in surface treatments for decreasing adhesion forces, for example, in the context of diamagnetically levitating \textmu m-scale or even smaller particles \cite{marti_pra,vienna_levitation,waarde_thesis}, or other experiments with miroparticles \cite{Ge2018density, Ge2020MagLev, Voisin24_PRL}.

Data underlying the results presented in this paper are available in the open-access Zenodo database: \href{https://doi.org/10.5281/zenodo.16902268}{10.5281/zenodo.16902268}.

\section*{Acknowledgments}
We thank Mats Hulander, Andreas Dahlin, and Romain Bordes for stimulating discussions about surface adhesion. We thank Anastasiia Ciers, Achintya Paradkar, Paul Nicaise, and Karim Dakroury for support with sample microfabrication and analysis, August Yurgens for fruitful feedback, and Thomas Penny for critical reading of the manuscript. This work was supported in part by the Horizon Europe 2021-2027 framework program of the European Union under Grant Agreement No. 101080143 (SuperMeQ), the European Research Council under Grant No. 101087847 (ERC Consolidator SuperQLev), the Knut and Alice Wallenberg (KAW) Foundation through a Wallenberg Academy Fellowship and Scholar (WW). Samples were fabricated at Chalmers Myfab Nanofabrication Laboratory.






\appendix

\begin{figure*}[b!htp]
    \centering
    \includegraphics[width=0.8\linewidth]{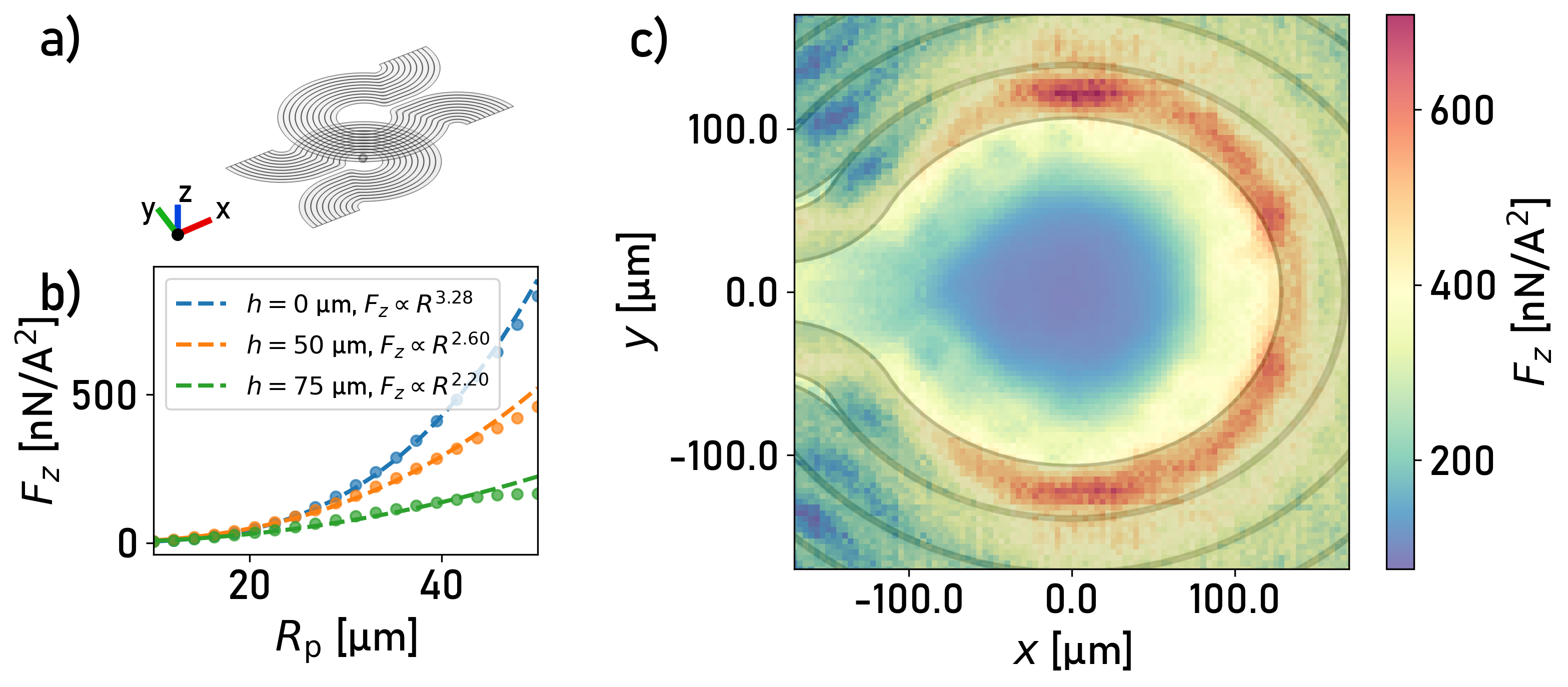}
    \caption{(a) The coil geometry used in the FEM simulation of the magnetic fields. (b) Lifting force at the centre as a function of particle radius, at different vertical planes of the trap. (c) Lifting force for a particle located at the plane of the bottom coil, with the outline of the coil shown.}
    \label{fig:tct_lift}
\end{figure*}

\section{Force calculations} \label{app:forces}

\subsection{Surface force and energy modelled via Lennard-Jones potential}
\label{app:ljpotential}

A well-used model for the force resulting from interacting surfaces is given by the Lennard-Jones potential \cite{lj_pot, lj_adhesion} of 8-2 type, which between infinitesimal two flat surfaces generates a surface energy
\begin{equation}\label{eq:lennard-jones}
    dE_S=-\frac{A_HdA}{2\pi z_{\text{eq}}^2}\left[\left(\frac{z_{\text{eq}}}{z}\right)^2-\left(\frac{z_{\text{eq}}}{z}\right)^8\right],
\end{equation}
where $A_H$ is the material specific Hamaker constant and $z_{\text{eq}}$ is the equilibrium distance between the two surfaces. We consider the two limiting cases of the particle being a flat disk or perfectly spherical, both with a radius $R_p$. If the particle is a flat disk, we find the total surface energy by integrating the contributions over the surface of the disk as
\begin{equation}\label{eq:E_disk}
    E_{\text{disk}}=-\frac{A_HR_p^2}{2z_{\text{eq}}^2}\left[\left(\frac{z_{\text{eq}}}{z}\right)^2-\left(\frac{z_{\text{eq}}}{z}\right)^8\right].
\end{equation}
For a spherical particle, we must take into account that the height $z$ varies, and we find after integration that
\begin{equation}\label{eq:E_sphere}
    E_{\text{sphere}}\approx-\frac{A_HR_p}{2z_{\text{eq}}}\left[\left(\frac{z_0}{z}\right)-\frac{1}{7}\left(\frac{z_{\text{eq}}}{z}\right)^7\right]
\end{equation}
under the approximation $R_p\gg z_{\text{eq}}$, which is the case for the scale of particles we are considering.

The bound surface energy is found through minimisation of the surface energy. We obtain
\begin{equation}\label{app:eq:surface_energy}
    \Delta E_{\text{disk}}= -{\left(\frac{127}{256}\right)}^{1/3}\frac{A_HR_p^2}{2z_{\text{eq}}^2}\text{ and } \Delta E_{\text{sphere}}=-\frac{3A_HR_p}{7z_{\text{eq}}},
\end{equation}
respectively. By finding the maximum to the force corresponding to \eqref{eq:E_disk} and \eqref{eq:E_sphere}, we obtain $F_{\text{disk}}\approx6.11\Delta E_{\text{disk}}/z_{\text{eq}}$ and $F_{\text{sphere}}\approx1.16\Delta E_{\text{sphere}}/z_{\text{eq}}$ as the maximal adhesion force that the interaction generates. We highlight that the bound surface energy $\propto R_p^2$ for disks and $\propto R_p$ for spheres, which together with the scaling law of the magnetic force yields a lower bound of particle size for which direct loading into a magnetic trap is possible. Note also that based on this study, the values for the exponents of \eref{eq:lennard-jones} are not determinable, as their values change only the numerical prefactors to the energies in \eref{app:eq:surface_energy}. Consequently, the model is insensitive to the exact adhesion mechanism.

\subsection{Magnetic force estimation} \label{app:magnetic_force_est}
For a planar circular coil of radius $R_{\text{coil}}$ placed at $z=0$ with large diameter compared to the wire width, the magnetic field along the central axis can be approximated by $\mathbf{B}(z)=B_0R_{\text{coil}}^3/{((R_{\text{coil}}^2+z^2)}^{3/2})\hat{z}$. A perfectly diamagnetic object with spatial extent $V$ is subject to a magnetic force, which for a pointlike object can be approximated as $\mathbf{F}=\frac{V}{\mu_0}\nabla|\mathbf{B}|^2$. It can be shown that such a coil will maximally generate a lifting force of $\mathbf{F}_{\text{lift}}=\pi B_0^2R_{\text{p}}^3\hat{z}/\mu_0 R_{\text{coil}}$, for a spherical perfect diamagnet of radius $R_{\text{p}}$ located $R_{\text{coil}}/2$ over the plane of the coil. While this estimate is rather crude, it reveals that we should expect lifting forces to be proportional to the volume of a diamagnetic object, that coils should be made as close as possible in size to the objects that one aims at levitating, and that forces on superconducting objects are limited by the critical field $B_C$ at which the superconducting state is destroyed.

For quantitatively accurate results, we employ FEM-based simulation to predict the lifting force and levitation frequencies for chip-based traps with realistic geometries (see \fref{fig:tct_lift}(a)), identical to the devices employed in Ref.~\cite{marti_pra}. We consider here the case of particle dimensions being much larger than the London penetration depth, approximating the superconductors as perfect diamagnets. Our procedure consists of first simulating the trapping coils without the presence of a particle, and then calculating the changes in the magnetic field due to its presence using the method employed in Ref.~\cite{hofer_sim} for a quadrupole field. A detailed description of this procedure is being prepared in another manuscript \cite{HOTs}. We find that regardless of trap current the lift force at the center of the bottom plane of a trap with the geometry shown in \fref{fig:tct_lift}(a) is maximally 11.7\,nN$\cdot\left(\frac{B_C}{1 \text{mT}}\right)^2$. In practice the force is also strongly dependent on particle position, see \fref{fig:tct_lift}(b) and \fref{fig:tct_lift}(c).

\begin{figure*}[t!hbp]
    \centering
    \includegraphics[width=\linewidth]{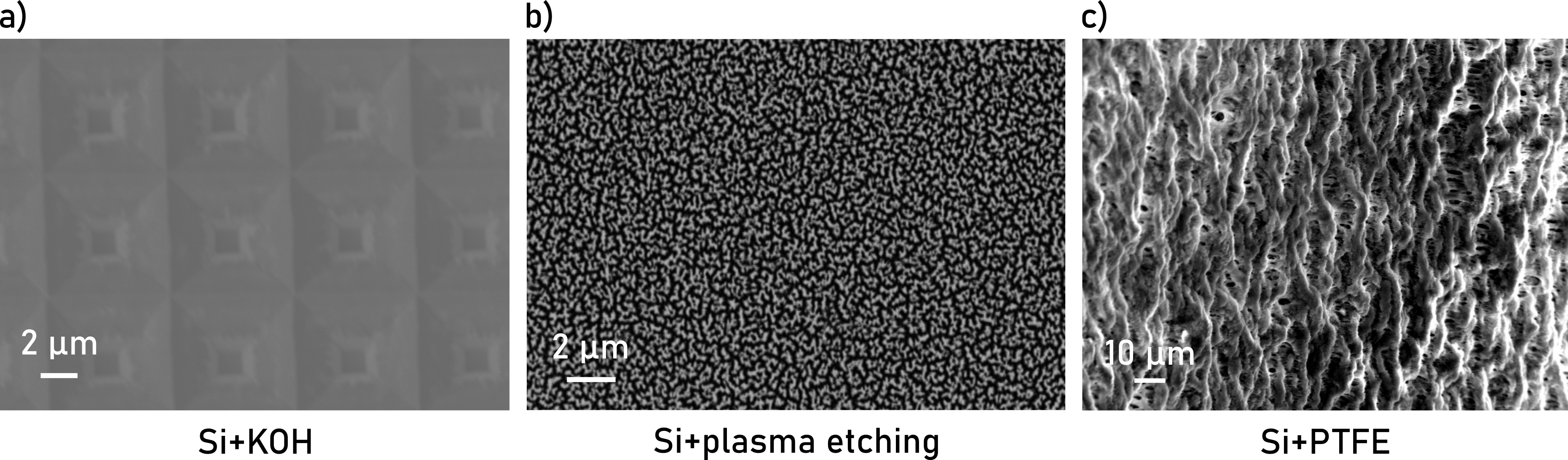}
    \caption{SEM images of the structured Si substrate after (a) KOH etching, (b) plasma etching, and (c) application of a PTFE membrane.}
    \label{fig:sem_all}
\end{figure*}

\section{Sample preparation}\label{app:sample_prep}

\subsection{Particles}

The particles used in the study consisted of an Sn63Pb37 alloy (EasySpheres), and were of mean diameter \SI{50}{\micro\meter} with a standard deviation of \SI{0.3}{\micro\meter}. The particles were stored in a vial of ethanol before the experiments. We have previously shown that the particles become superconducting below \SI{6.4}{\kelvin} with a critical field $B_{C1}(0)=\SI{65}{\milli\tesla}$ based on susceptibility measurements \cite{marti_ieee}. 

Particles were taken from the vial, dispersed on a transfer wafer, dried and then moved using a micromanipulator needle to the surface under test. Note that suspending the particles in solvent is not absolutely necessary and particles could also be transferred completely dried. We observed that directly dispersing the particles from solvent onto the surface under test leaves a visible solvent residue on the surface and makes the adhesion stronger. Thus, it is key that the particles are dried and individually transferred not to have solvent contamination influencing the result.

\subsection{Si substrate preparation}

After sample processing, the protective resist used during separation of individual chips from the wafers was stripped and the chips were ultrasonically cleaned in IPA, washed with water and dried using compressed nitrogen. The chips were then glued with BF-6 glue to glass slides made to fit the experimental apparatus. We expect the surface energy to be about \SI{2}{\joule\per\meter\squared} \cite{Jaccodine63_SiSurfaceEnergy}.

\subsection{Microstructuring of Si (KOH etching)}

Microstructuring of the silicon surface was performed using KOH etching, which etches $\langle 100\rangle$ cut silicon anisotropically with an etch angle of 57\,\degree. A 50\,nm layer of SiO$_2$ was deposited by magnetron sputtering on the chips to act as a hardmask for the etching. Designs of square hole lattices of different sizes were prepared and transferred onto the chips using optical lithography. The exposed SiO$_2$ was then dry-etched using reactive-ion-etching with flourine chemistry. The resist was stripped and KOH etching was performed in a 35\% solution at 80 \degree C. After etching, the hardmask was removed in a 2\% hydrofluoric acid bath. The result of this fabrication process was well defined quadratic inverse pyramids, as can be seen in \fref{fig:sem_all}(a). The base dimension was lithographically defined to be \SI{7}{\micro\meter} and the etching depth was measured with profilometry to be \SI{1.75}{\micro\meter}. For this kind of microstructuring (feature sizes $\sim 5$\,\textmu m), a slight decrease in water contact angle was noted compared to the unstructured surfaces, which can be explained by the Wenzel model \cite{Wenzel1936}.

\subsection{Plasma nanostructuring of Si (black silicon)}

By careful tuning of parameters of an RIE process, surfaces of Si samples can be structured to form free-standing pillars of nm-size. The light scattering properties of such a surface will make it appear dark compared to unstructured Si, hence the term black silicon \cite{black_si95}.

We produced black-looking surfaces by etching unpatterned Si chips in an ICP-RIE system (Oxford PlasmaPro 100) with an O$_2$:SF$_6$ flow ratio of 3:1, CCP power of 100 W and ICP power of 500 W at 50 mTorr of chamber pressure. Etching was continued until the surfaces no longer were growing darker in color, which for a 2 inch wafer was about 60 minutes.

An AFM scan was made of the surface of the plasmastructured samples, the result being shown in \fref{fig:afm_collage}(b). Additionally, an SEM image of the sample is shown in \fref{fig:sem_all}(b). We were not successful in achieving sharp pillar-like features, but still managed to produce a surface with high nanometer-scale roughness.

\subsection{PTFE membranes}

PTFE membranes were prepared from standard thread sealant tape by carefully cutting it into appropriately sized pieces, submerging them in IPA solvent and placing them onto the chip surfaces. The samples were dried on a hot plate in order to evaporate off the solvent, yielding a well bonded membrane with no visible air bubbles. The resulting membrane was rather soft and had a thickness measured to be 75\,\textmu m with an optical profilometer. A SEM picture of the membrane on top of a sample is shown in \fref{fig:sem_all}(c). From this image and the AFM scans that were captured, we ascertain that the applied membrane consists of a porous structure, resulting in surface roughness simultaneously on the 10\,nm and 100\,nm scale. A fibrous structure is apparent, which is presumably a result of uniaxial stretching performed during the manufacturing process of the tape \cite{ptfe_handbook}. The surface free energy is expected to be about \SI{20}{\milli\joule\per\meter\squared} \cite{Kinloch12_adhesion}. From the AFM scans, an RMS roughness of 87\,nm was calculated, although the scanned area would have to be larger to capture the full roughness of the surface.

We observed that during particle placement, the soft nature of the membrane sometimes resulted in the particles deforming the membrane and being partially encapsulated by the material. This results in increased area of contact, and we observed that this increased the force required to release the particles from the surface. It is therefore essential with soft coatings to carefully detach the particles from the transfer device onto the layer. All samples used in the study were prepared in this way.

\subsection{Parylene polymer}

Coatings of parylene polymer were deposited using a CVD system (SCS PDS 2010 Labcoter) onto the Si surfaces. The monomer used was Parylene C, with a target thickness was 200 nm. Some of the samples were then exposed to 25 W of O$_2$ plasma for 60 s and 50 W of CF$_4$ plasma for 30 s with the intention of creating a roughened surface terminated with fluorine groups, with a high water contact angle \cite{parylene_o2_plasma} and consequently small adhesion to the particles. The WCA was about 85\degree{} for the samples before plasma treatment and above 100\degree{} for the treated samples. The surface energy of the parylene coating is expected to be around \SI{40}{\milli\joule\per\meter\squared} \cite{Golda-Cepa14_parylene}.

\subsection{PMMA coating}
A widely accessible coating material is poly(methyl methacrylate) (PMMA), in the spin-coatable form used as e-beam resist. We spin-coated onto Si chips 950 PMMA A6 at 3000 rpm and 50 s at an acceleration of 500 rpm/s, baked at 180\degree{} for 2 minutes and measured the effects. This resulted in a uniform layer of about 600 nm of thickness. An increase in water contact angle compared to Si to about 90\degree{} was observed.

Like parylene, the surface of PMMA is known to be possible to nanostructure using O$_2$ plasma \cite{o2_pmma_sturcturing}, which could serve to further decrease surface adhesion. Fluorination of the film is also possible through subjecting the films to plasmas containing fluorine radicals \cite{pmma_fluorination}. We attempted to mimic this by subjecting PMMA coated Si chips to plasmas containing fluorine chemistry, but failed to see any beneficial effects in the adhesion properties, and present results for untreated PMMA coatings. The surface free energy of PMMA is expected to be around \SI{40}{\milli\joule\per\meter\squared} \cite{Kinloch12_adhesion}.

An AFM scan was made of the surface, the result being shown in \fref{fig:afm_collage}(e). It was found that the surface was only very slightly rougher than the Si substrate, having a mean surface roughness of $S_q=\SI{0.4}{\nano\meter}$.

\subsection{HSQ coating}
Hydrogen silsesquioxane (HSQ) is another polymer widely available as e-beam resist. The resist was purchased as a dry powder (EM Resist Ltd), mixed to a 6 \% ratio by weight with methyl isobutyl ketone (MIBK) in an oven-dried bottle. Before application onto the Si substrates by spin coating, the solution was filtered through a syringe strainer.

The solution was spin coated onto the chips at 3000 rpm and baked at 110 degrees for 1 minute. The resulting layer had a thickness of 135 nm, measured using profilometry, and a water contact angle of 85\degree. In order to exclude the possibility that the MIBK solvent had any effects on the surface, we applied pure MIBK to the substrates in the same way. We saw no effect on either the water contact angle or the appearance of the surfaces, and thus concluded that the effects seen were due to the presence of the HSQ.

\subsection{Au coating by sputtering}
Using a DC sputtering tool (FHR MS 150) a layer of Au of approximately 75 nm of thickness was sputtered on Si substrates. Under the Au layer, a Ti layer was used for the layer to adhere well onto the substrates. We found that the resulting thin film was very susceptible to scratches during handling, and thus made sure to place the particles on pristine regions on the chips. The surface energy of the Au layer is expected to be around \SI{750}{\milli\joule\per\meter\squared} \cite{Tran2016_surface_energy}.

In the case that electrostatic forces are the origin of the adhesion between the particles and substrates, it is expected that the charges should equilibrate if both the particles and substrates are conducting, which served as motivation for this preparation. In a chip-based magnetic trapping device, having normal conductors of the vicinity of the levitation region will induce dissipation through eddy currents \cite{twamley_eddy, fuwa_ferromagnet}, which may limit the quality factor of the motion and are thus undesirable.

\begin{figure*}[t!hbp]
    \centering
    \includegraphics[width=0.8\linewidth]{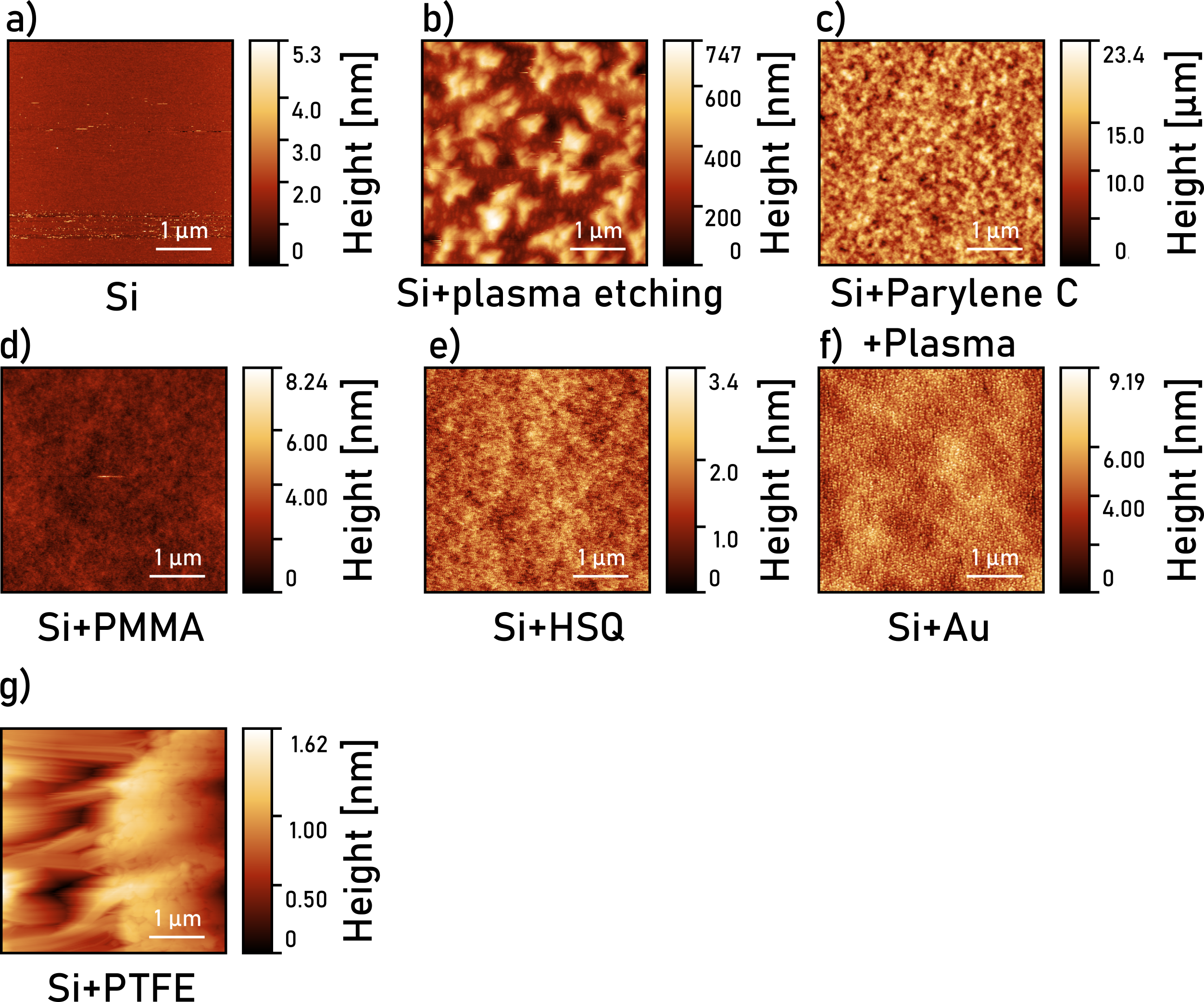}
    \caption{Scans of the sample height taken with AFM in peak-force mode. The determined mean surface roughness and mean adhesion force are found in \tref{tab:stats}.}
    \label{fig:afm_collage}
\end{figure*}

\section{Complete measurement data}\label{app:data}

\subsection{Atomic force microscopy scans}\label{app:afm}

\fref{fig:afm_collage} shows AFM scans of the samples over a 4\,\textmu m x 4\,\textmu m scan area.

\subsection{Water contact angle measurements}\label{app:WCA}

\begin{figure*}[t!hbp]
    \centering
    \includegraphics[width=0.6\linewidth]{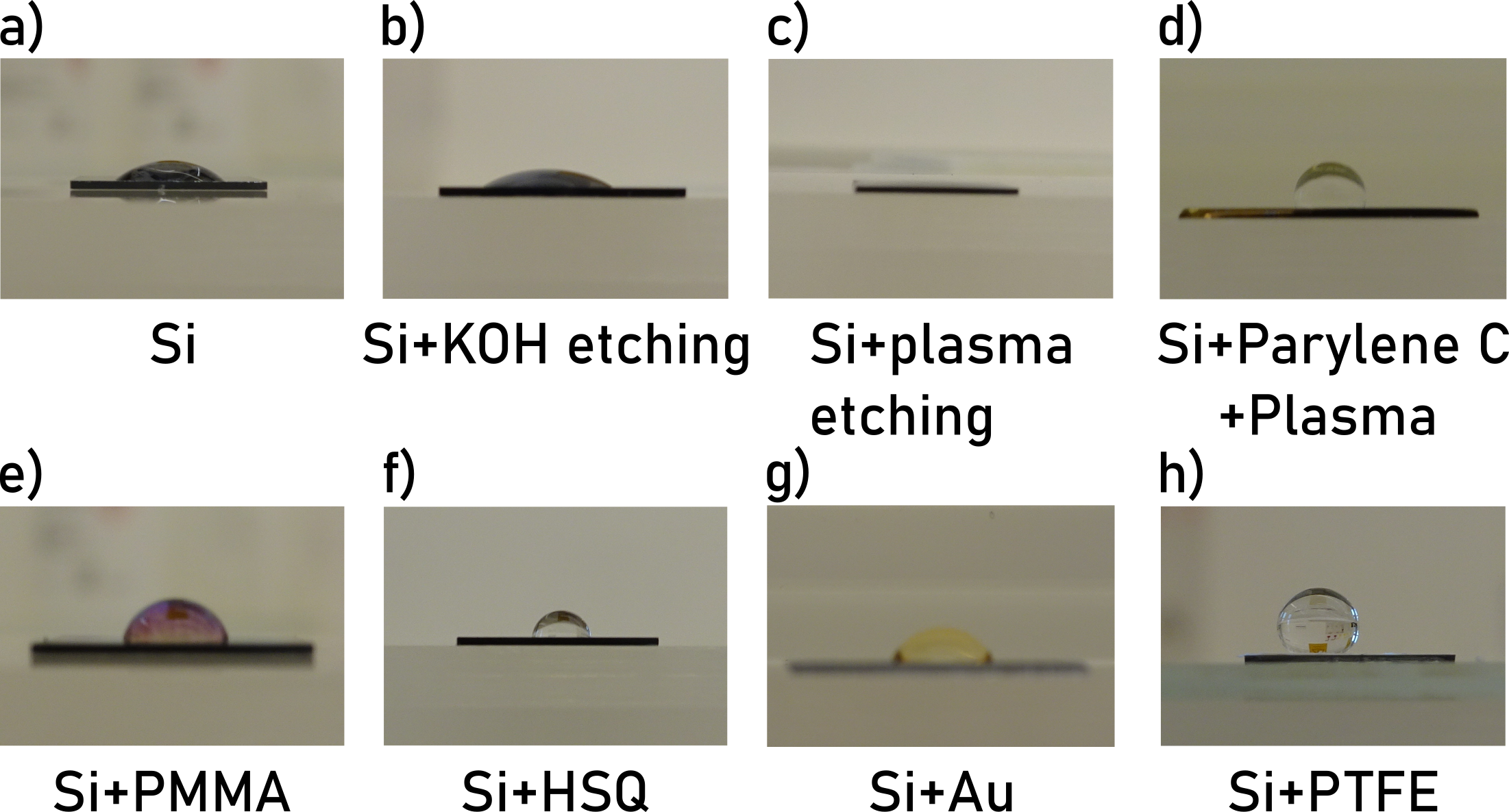}
    \caption{Images of water droplets on top of the sample surface for all the surface preparations used in our work. We determine the water contact angle from these images, see  \tref{tab:stats}.}
    \label{fig:wca_collage}
\end{figure*}

\fref{fig:wca_collage} shows images of samples with water droplets used to determine the WCA. Deionized water was applied with a micropipette, and a digital camera was used to capture the photos. The WCA was then determined by manually analysing the images in photo editing software.

\subsection{Detachment force measurements and parameter estimation}\label{sec:all_detachment_cdf}

\begin{figure*}[t!hbp]
    \centering
    \includegraphics[width=\linewidth]{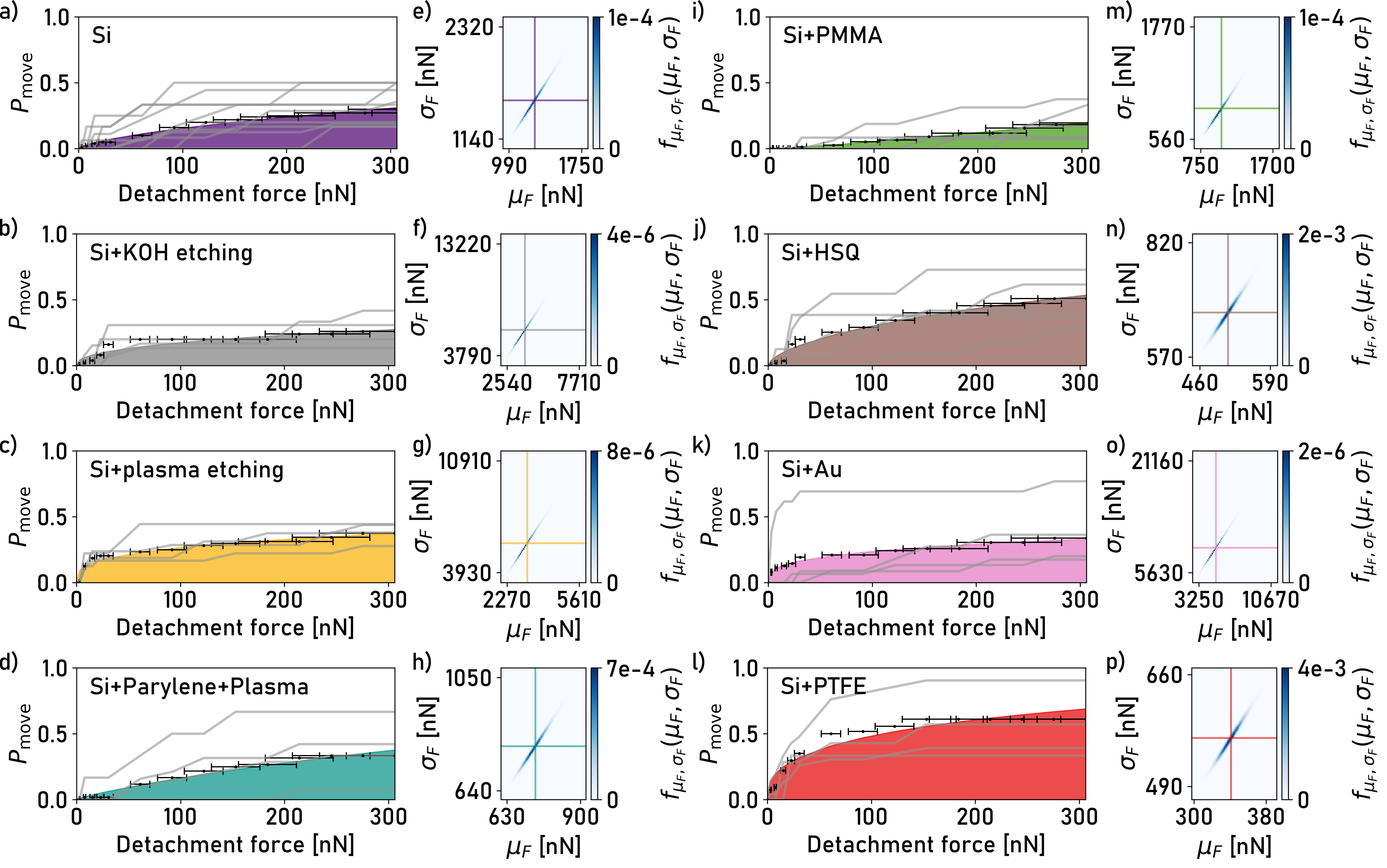}
    \caption{Cumulative distribution functions (a-d, i-l) for the engineered surfaces along with associated histograms (e-h, m-p) of the probability distributions of the parameters to the gamma distribution.}
    \label{fig:cdf_collage}
\end{figure*}

We show in \fref{fig:cdf_collage} the measured cumulative distribution functions for all samples used in the study, as well as the model fits.

We use a gamma distribution model to obtain a statistical model for the detachment probability with Bayesian parameter estimation \cite{Gregory_2005}. We use uniform priors, and sample the posterior distribution using MCMC with the Metropolis-Hastings algorithm through the \texttt{emcee}-library \cite{emcee}. For the likelihood we use a Gaussian, such that
\begin{equation}
    f(D|\mu_F, \sigma_F)\propto\exp{\left(-\sum_{i}\left[\frac{P_i-P_\Gamma(F_i|\mu_F, \sigma_F)}{\sigma}\right]^2\right)},
\end{equation}
where $D$ denotes the data set composed of the forces $\{F_i\}$ and corresponding moving probabilities $\{P_i\}$. The prediction from the model $P_\Gamma(F_i|\mu_F, \sigma_F)$ is the cumulative distribution function of the gamma distribution, which parametrised with mean and standard deviation is
\begin{equation}
    P_\Gamma(F_i|\mu_F, \sigma_F)=\frac{\gamma(\mu_F^2/\sigma_F^2, \mu_FF_i/\sigma_F^2)}{\Gamma((\mu_F/\sigma_F)^2)}.
\end{equation}
where $\gamma(s, x)$ is the lower incomplete gamma function and $\Gamma(x)$ is the $\Gamma$-function. By $\sigma$ we denote a particle miscounting error, which we set to 1 \%.

The samples allow for producing the histograms in \fref{fig:si_cdf} and \fref{fig:cdf_collage} of the joint posterior probability distribution function $f_{\mu_F, \sigma_F}(\mu_F, \sigma_F)$, as well as obtaining the 16th, 50th, and 84th percentiles of the data. We take into account the uncertainty in the calibration in a maximal error sense, by calculating the 84th percentile with the largest calibration factor and the 16th with the smallest. For the 50th percentile we take the mean of the calibration factors. Simple root finding can now be used to find $F_{P=0.5}$, the force at which 50\% of particles are predicted to move using the estimated parameters. We present the results of the parameter estimation in \tref{tab:stats}, noting also that the mean and standard deviation are correlated parameters for the gamma distribution.





\begin{figure*}[t!hbp]
    \centering
    \includegraphics[width=0.5\linewidth]{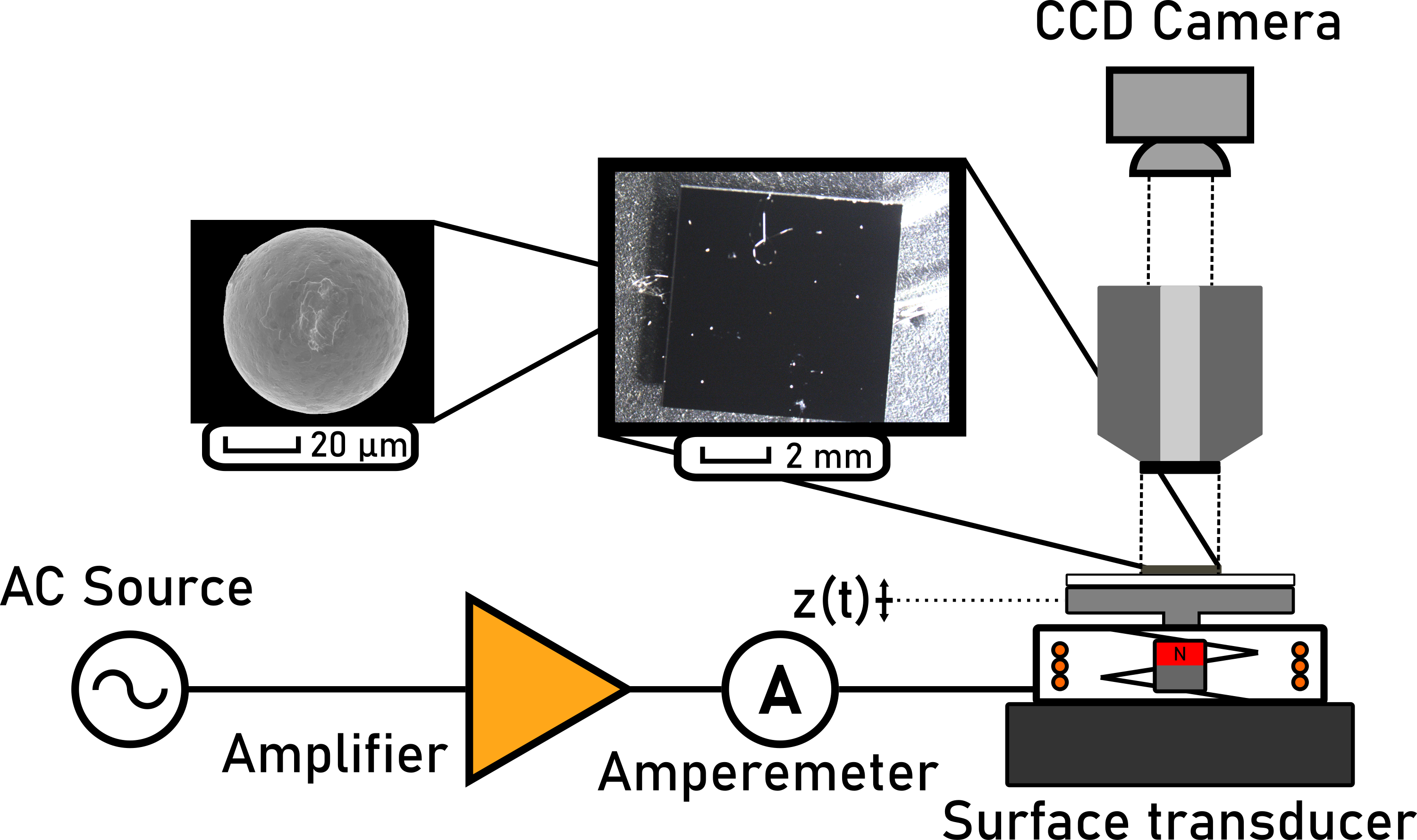}
    \caption{Sketch of the experimental setup to measure detachment force of particles placed on a sample. The top of the surface transducer can be fit either with a MEMS accelerometer for calibration or with a sample chip for actual measurements.}
    \label{fig:shaker}
\end{figure*}

\section{Shaking experiment}\label{app:setup}

\subsection{Experimental setup}

\fref{fig:shaker} shows the experimental setup used to determine the detachment force of particles from sample surfaces. Substrates with the surfaces under test were glued using BF-6 glue to microscope glass slides, which were attached to a surface transducer. A sinusoidal excitation signal amplified with a standard audio amplifier was applied to the transducer in order to induce vibrations to the samples. By choosing to work at acoustic frequencies, commercially available accelerometers are fast enough to be able to calibrate the induced accelerations and no high-voltage electronics are required. Using a MEMS accelerometer (Kioxia KXI134-1211), the applied displacement signal was calibrated to an applied acceleration, which together with the mass of the particles were used to calculate the force applied to the particles. The apparatus was able to generate accelerations above the $\pm$64$g$ range of the accelerometer, but based on measurement of the current consumption of the amplifier we estimate that this acceleration is close to saturation of the system. All presented measurements shown from this experiment are within the calibrated range. In units of force, the maximum generated by the device corresponds to 306\,nN, comparable to that of the superconducting chip trap devices presented in previous works \cite{marti_ieee,marti_pra}.

The experiment was observed with a camera through a microscope lens, allowing for taking images of the samples as acceleration is applied to the substrates. The sample was illuminated using white LEDs, with the magnification of the microscope adjusted to have the entirety of the sample in frame.

\begin{figure*}[t!hbp]
    \centering
    \includegraphics[width=\linewidth]{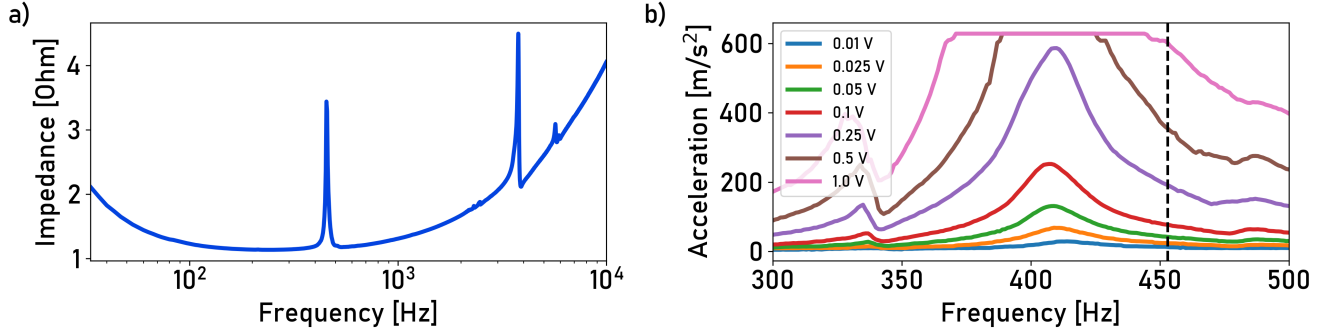}
    \caption{(a) The combined line impedance of the surface transducer and audio amplifier, within the bandwidth of the audio amplifier taken at an excitation voltage of 0.1 V. (b) Accelerations measured using the MEMS accelerometer for frequencies in the range 300\,Hz to 500\,Hz. The clipping of the curves is due to the accelerometer reaching its maximum output value.}
    \label{fig:impedance}
\end{figure*}

\begin{figure*}[t!hbp]
    \centering
    \includegraphics[width=\linewidth]{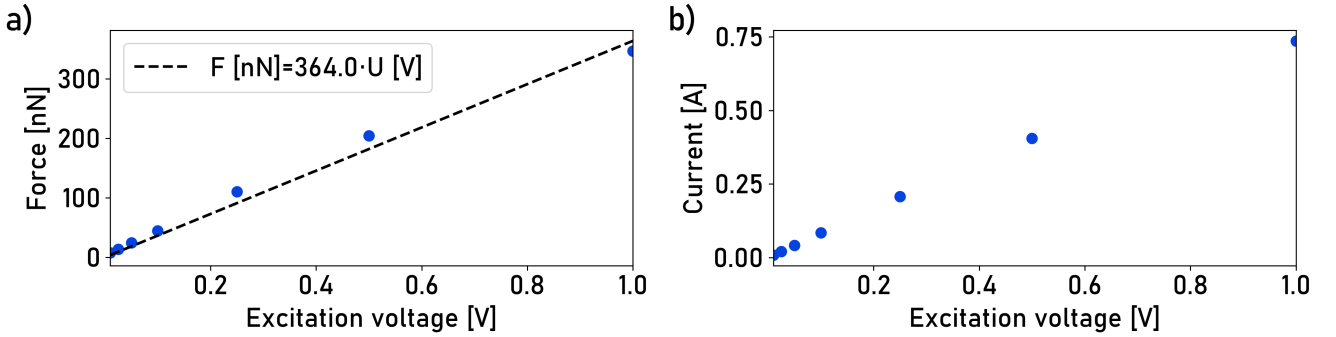}
    \caption{(a) The force exerted on 50 \textmu m particles by the device, with a linear fit to the force. (b) Measured current through the surface transducer per applied voltage is shown in the calibrated range.}
    \label{fig:acc_fit}
\end{figure*}

\subsection{Calibration}\label{app:mems_calibration}

The impedance of the surface transducer was measured using an AC amperemeter connected after the amplifier and sweeping the excitation frequency. \fref{fig:impedance}(a) shows the measurement result. Two clear resonances are found in the pass band at $\sim$410\,Hz and $\sim$3500\,Hz. The working frequency of the experiment ($f=453$\,Hz) was chosen to be slightly above the lower resonance frequency in order to have an acceleration signal in the bandwidth of the accelerometer, and have an acceleration response function insensitive to non-linear frequency shifts at high amplitudes.

Using a RP2040 microcontroller, the acceleration of the surface transducer was read out over the SPI bus from a MEMS accelerometer (Kioxia KXI134-1211). With this, it was possible to read out accelerations at a range of $\pm 64 g$ at signed 16 bit precision with a sample rate of roughly 4000 Sa/s.

The acceleration transfer function was then measured by sweeping the frequency and excitation strength in a region around the working frequency. The data was used to produce plots of the type shown in  \fref{fig:impedance}(b), to which a polynomial function could be fitted describing the relationship between acceleration and excitation voltage. The result is shown in \fref{fig:acc_fit}. In practice, the experiments are restricted to voltages up to 1V, for which the response acceleration to good approximation is a linear function. The calibration procedure was performed several times over the course of the experiment, and it was found that the resonance frequency had a temporal drift. This in turn caused the calibrated force per voltage to vary between 250-364 nN/V over the course of the experiment. The error intervals in \tref{tab:stats} are corrected to include the maximal error that the resonance drift may result in, and other measurements are presented with the mean of the calibration factors.

\begin{figure*}[t!hbp]
    \centering
    \includegraphics[width=0.8\linewidth]{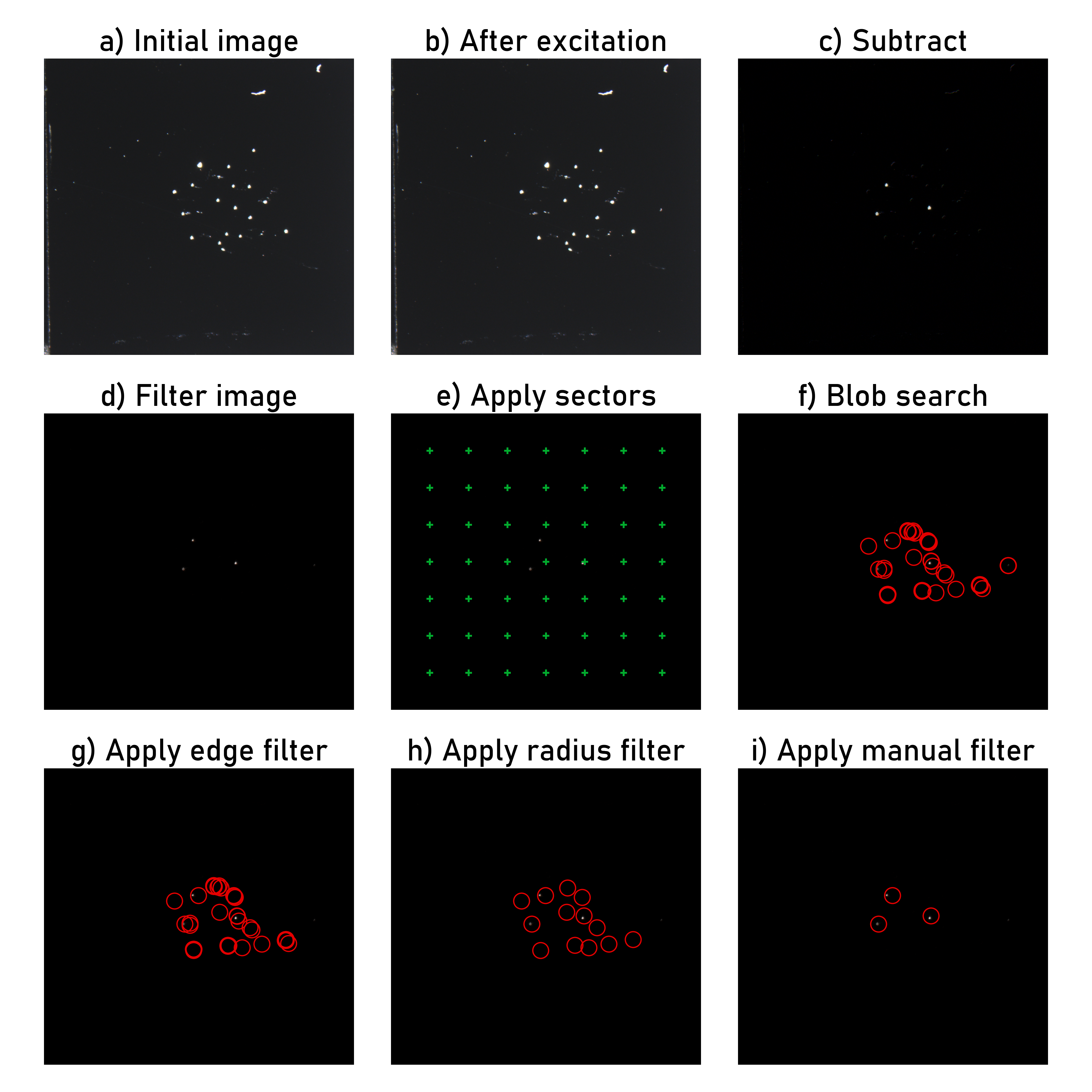}
    \caption{Graphical description of the steps of the search algorithm for moving particles. }
    \label{fig:particle_search}
\end{figure*}

\subsection{Detection algorithm for particle movement}\label{app:filtering}
While powerful AI-driven image analysis tools that would be able to detect particles having moved between two images are available, we employ a simpler method to identify particles. Initially, the image is rotated to align the chip axis with the image axis . The image is then cropped to the chip edges using the pixel brightness gradient having a maximum at the transition between chip and background (see \fref{fig:particle_search}(a)).

The identification process then starts with subtracting the pixel color values of the original image with the color values of an image of the sample taken after applying a given detachment force (\fref{fig:particle_search}(b)) to the sample. The resulting image will consist of bright sports where objects have moved on a darker background (\fref{fig:particle_search}(c)). To remove the residual noise on the dark background, the image is subjected to a brightness filter such that
\begin{widetext}
\begin{equation}
    (R',G',B')_{i,j}=\begin{cases}(R,G,B)_{i,j}\text{ if mean}[(R,G,B)_{i,j}] > t\cdot\text{mean}_{i,j}\left(\text{mean}[(R,G,B)_{i,j}]\right)\\
    0\text{ otherwise},\end{cases}
\end{equation}
\end{widetext}
where $t$ is a threshold parameter to the filter. As next step, the image data is subjected sequentially to a minimum filter, a median filter and then another minimum filter. This removes most of the noise from the image while keeping the brighter blobs due to moving objects and prepares for the searching algorithm (\fref{fig:particle_search}(d)).

In order to increase the efficiency of the search, the image is divided into sectors (\fref{fig:particle_search}(e)) with every sector only being searched if the mean brightness of the sector exceeds the mean brightness of the image as a whole. With the image divided into an appropriate amount of sectors, this can significantly increase the speed of the algorithm. Next, if the sector is accepted, it is filtered with a window that sets all pixel values inside the window white if the number of pixels with brightness larger than a threshold exceeds another threshold number. This enlarges the size of the bright features of the sector. After this the sector is gaussian blurred and finally submitted to the \texttt{blob log}-function as implemented in the \texttt{scikit-image}-library \cite{scikit-learn} (\fref{fig:particle_search}(f)). The resulting identified locations are then tentatively taken as moving particles.

Care was taken to place the test particles in the central part of the sample, which meant that some falsely identified locations could be immediately rejected due to their proximity to the sample edge (\fref{fig:particle_search}(g)). In most samples, dirt or imperfections in the surface were present which were falsely identified as particles. While care was taken to physically avoid this happening, the movement counting script included several ways to filter out unwanted counts. When some types of dirt was present on the sample chips and moved, the script would wrongfully identify it as a particle. In order to reject these locations, an adjacency clause was implemented which for a preset radius of particle $R$ rejected moving particles that had a chain of overlaps greater than another preset parameter (\fref{fig:particle_search}(h)). Specific samples often required tuning of the filter parameters in order to efficiently reject all false positives, while not rejecting any true positives. Instead, the algorithm was deliberately tuned to be admissive, and remaining false identified locations could be manually rejected by inspection of the initial images (\fref{fig:particle_search}(i)).

The initial number of particles on the chips were counted manually before each experimental run, and all particles counted as moving were manually checked against the unprocessed images.

\clearpage


\bibliography{referenser}

\end{document}